# A Theoretical Basis for Determination of the Reynolds Stress in Canonical Turbulent Flows

T.-W. Lee* and Jung Eun Park
*Mechanical and Aerospace Engineering, Arizona State University, Tempe, AZ, 85287*

**Abstract- We present a unique method for solving for the Reynolds stress in turbulent canonical flows, which is based on momentum balance for a control volume moving at the local mean velocity. Comparisons with experimental and computational data in simple geometries show quite good agreement. An alternate picture for the turbulence momentum transport is offered, based on the closed-form expression for the Reynolds stress. The turbulence momentum balance is verified using DNS data. This approach has potential applications in aerodynamics, and also represents a possible method for determining the Reynolds stress in more complex flows.**

*Corresponding author:
T.-W. Lee
Mechanical and Aerospace Engineering, SEMTE
Arizona State University
Tempe, AZ 85287-6106
Email: attwl@asu.edu

## Nomenclature

| | | |
|---|---|---|
| $C_1$, $c$ | = | constants |
| $Re_\tau$ | = | Reynolds number based on friction velocity |
| $U$ | = | mean velocity in the x direction |
| $U_e$ | = | free-stream velocity |
| $V$ | = | mean velocity in the y direction |
| $u'$ | = | fluctuation velocity in the x direction |
| $u_{rms}'$ | = | root-mean square of u' |
| $u'v'$ | = | Reynolds stress |
| $\delta$ | = | boundary layer thickness |
| $\nu_m$ | = | modified kinematic viscosity |

## I. Introduction

Turbulence is considered one of the most difficult problems in fluid physics, or in physics in general. It is also quite important as many issues of practical concern, such as aerodynamics, combustion flows, meteorology and many industrial processes, depend on turbulence, and much work has been done on finding some adequate approximations so that immediate problems of turbulent flows can be solved through turbulence "models" (we do not attempt to list the vast literature in this area).

As finding the entire absolute (mean + fluctuations) velocity field is quite difficult, or as some argue an overflow of information, here we use a formulation to find the Reynolds stress in terms of "root" turbulence variables, such as mean velocity, and longitudinal kinetic energy. This kind of analytical approaches for closure have been few and far between, whereas there has been no shortage of developments in approximate methods such as turbulence models or large-eddy simulation. Ambitious series of work by Kraichnan [1-3] attempts to cast the Navier-Stokes equation in a form that relates the interaction between eddies of different scales, so-called direct interaction approximation. In the end, however, a suggestion is made for a search for “Lagrangian closure” [3]. Although no significant advances have been made in that regard, to our knowledge, Lagrangian methods have been frequently used for analyses of mixing and dispersion [4]. Also, probability density function modeling of mixing relies on Lagrangian framework [5]. A low-order model of Reynolds stress has been developed by Egolf and co-workers [6, 7], where the Reynolds-averaged Navier-Stokes equation can be simplified and an expression for the Reynolds stress is found after applying the boundary conditions for simple flow geometries.

In this work, we attempt to demonstrate that an expression for the Reynolds stress can be found by simply using a control volume that is moving at the local mean velocity. In contrast to turbulence models where higher-order terms are constantly produced and need to modeled, as in k-ε models or Reynolds stress models, the current approach leads to decoupling of the variables so that the Reynolds stress is written in terms of “lower-order” terms. No ad-hoc modeling is required. In this work, we attempt to show how the integral formula can produce Reynolds stress which match the experimental and direct numerical simulation (DNS) data flows in canonical geometries.

## II. Integral Formula and Its Theoretical Formulation

The integral formula for the Reynolds stress is given in Eq. 1. In Eq. 1 and subsequently, we omit the typical bar notation above the variables that denote time averages. U is the mean streamwise velocity, while primes denote time-averaged fluctuation velocities. $C_1$, c, and $\nu_m$ are constants.

$$u'v' = -C_1\left[Uu'^2 - c\int_0^y \frac{dU}{dy}u'^2 dy\right] + \nu_m \frac{\partial u_{rms}'}{\partial y} \qquad (1)$$

Eq. 1 arises from the momentum balance for a Lagrangian control volume as depicted in Fig. 1. The derivation is included in the Appendix. Briefly, in Fig. 1 the momentum balance for a Lagrangian control volume moving the mean velocity:

$$\frac{\partial(u'v')}{\partial y} = -\frac{\partial u'^2}{\partial x} + \nu_m \frac{\partial^2 u_{rms}'}{\partial y^2} \qquad (2)$$

We note that for a moving control volume the mean velocity is decoupled from the fluctuating components since the observer is moving at the local mean velocity, after also making the boundary layer approximate ($V<<U$) and neglecting the laminar viscous term. In this sense, the momentum balance is Galilean-invariant. Thus, the change in the Reynolds stress in the transverse direction is mainly due to excess streamwise fluctuating momentum, plus the shear stresses due to instantaneous gradient in the fluctuating velocity, which is time-averaged (this quantity is *not* assumed to be zero). Pressure term is neglected. The remaining steps from Eq. 2 to 1 are shown in the Appendix.

Eq. 1 has two constants ($C_1$, with a dimension of inverse of velocity, and dimensionless c) that depend on the Reynolds number. $C_1$ is related to the geometry of the flow, and also boundary or symmetric conditions, and in some cases a geometrical relationship may be developed to give a direct expression for $C_1$ as a function of the Reynolds number and the flow configuration. $\nu_m$ is the modified laminar viscosity. Eq. 1 shows that the Reynolds stress at a given location (y) arises primarily from the excess turbulence transport ($Uu'^2$) in the streamwise direction, and is reduced by the cumulative displacement away from that location, represented by the integral term. $u'^2$ is of course also the diagonal component of the Reynolds stress. From Eq. 1, we can see that the mean velocity gradient (as in the so-called turbulence production) term appears, but *inside* an integral. The Boussinesq hypothesis relates the Reynolds stress, as $u'v' = \nu_T(dU/dy)$. The mean velocity gradient effect is still present in Eq. 1, but it is *cumulative*, and also appears indirectly in the first term on the RHS. This fact is a small hint (the numerical value of the integral term is small relative to other terms as shown later) that insofar as turbulence models are concerned, the Reynolds stress modeling may be preferable as it bypasses the Boussinesq hypothesis.

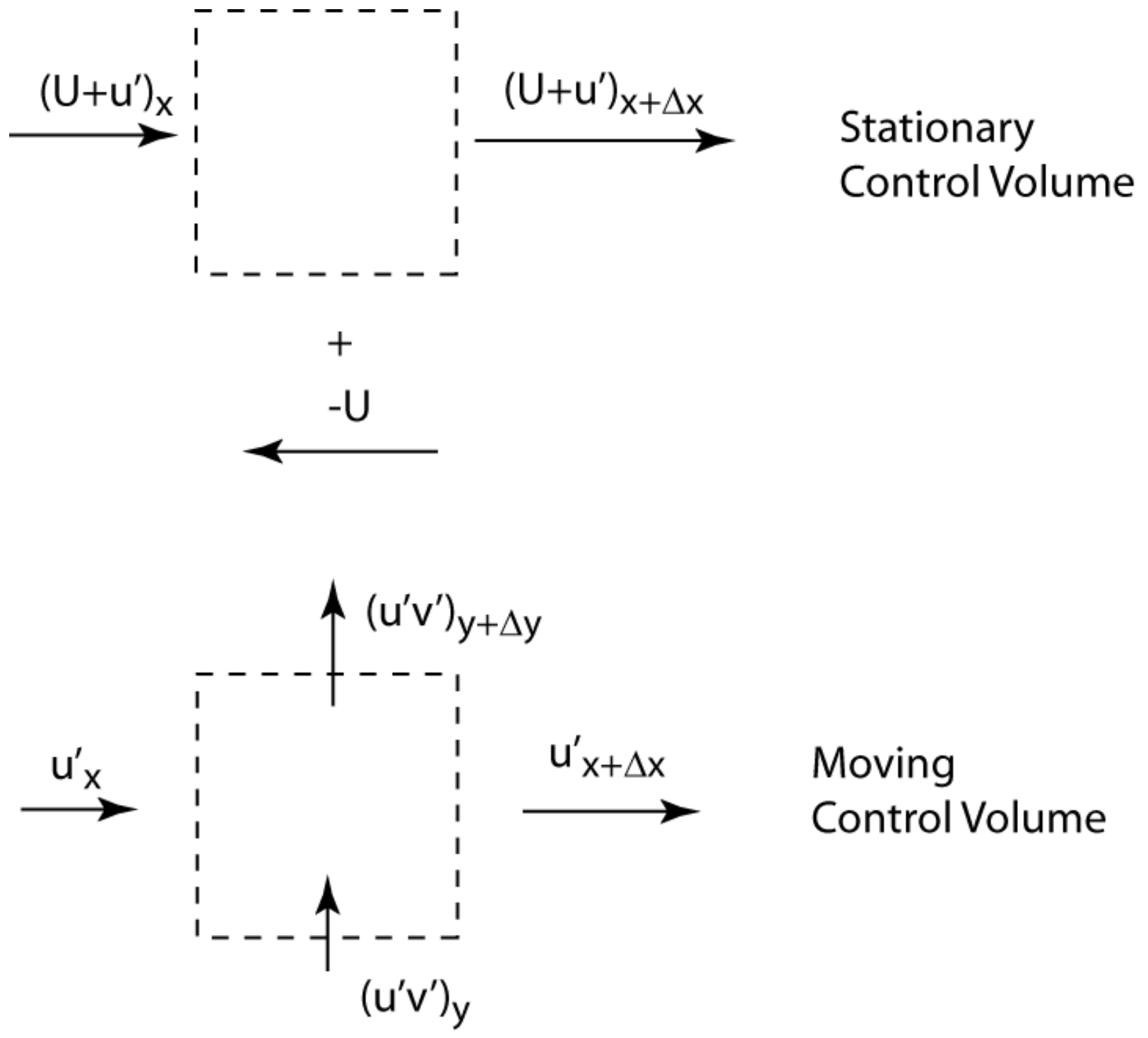

**Fig. 1. Illustration of the momentum balance for a moving control volume.**

## III. Comparison for Boudary-Layer Flow over a Flat Plate

Figure 2 shows the comparison of the Reynolds stress obtained from Eq. 1 with experimental data of DeGraaf and Eaton [8]. In that work, data on various turbulence quantities and the Reynolds stress (all normalized by the friction velocity) are provided, and also various scaling approaches tested with the data, in a well-designed experiment for flows over a flat plate with zero pressure gradient. The Reynolds number based on the momentum thickness ($Re_\theta$) ranged from 1430 to 31000 [8]. We use their measured parameters, U and $u'^2$, as given in DeGraaf and Eaton [8], and input them into Eq. 1. Gradients of U and $u'^2$ are calculated from the experimental data. As the experimental data are discontinuous, and at times hard to transcribe, there are some fluctuations and potential errors in the final calculations of the Reynolds stress, particularly close to the wall where the gradients are very steep and the data points all clustered. We can nonetheless input the root parameters into Eq. 1 to compute accordingly, and compare with the measured Reynolds stress as in Figure 2. In spite of dealing with discontinuous experimental data and their gradients, the comparison of Eq. 1 with experimentally observed Reynolds stress is in general quite good.

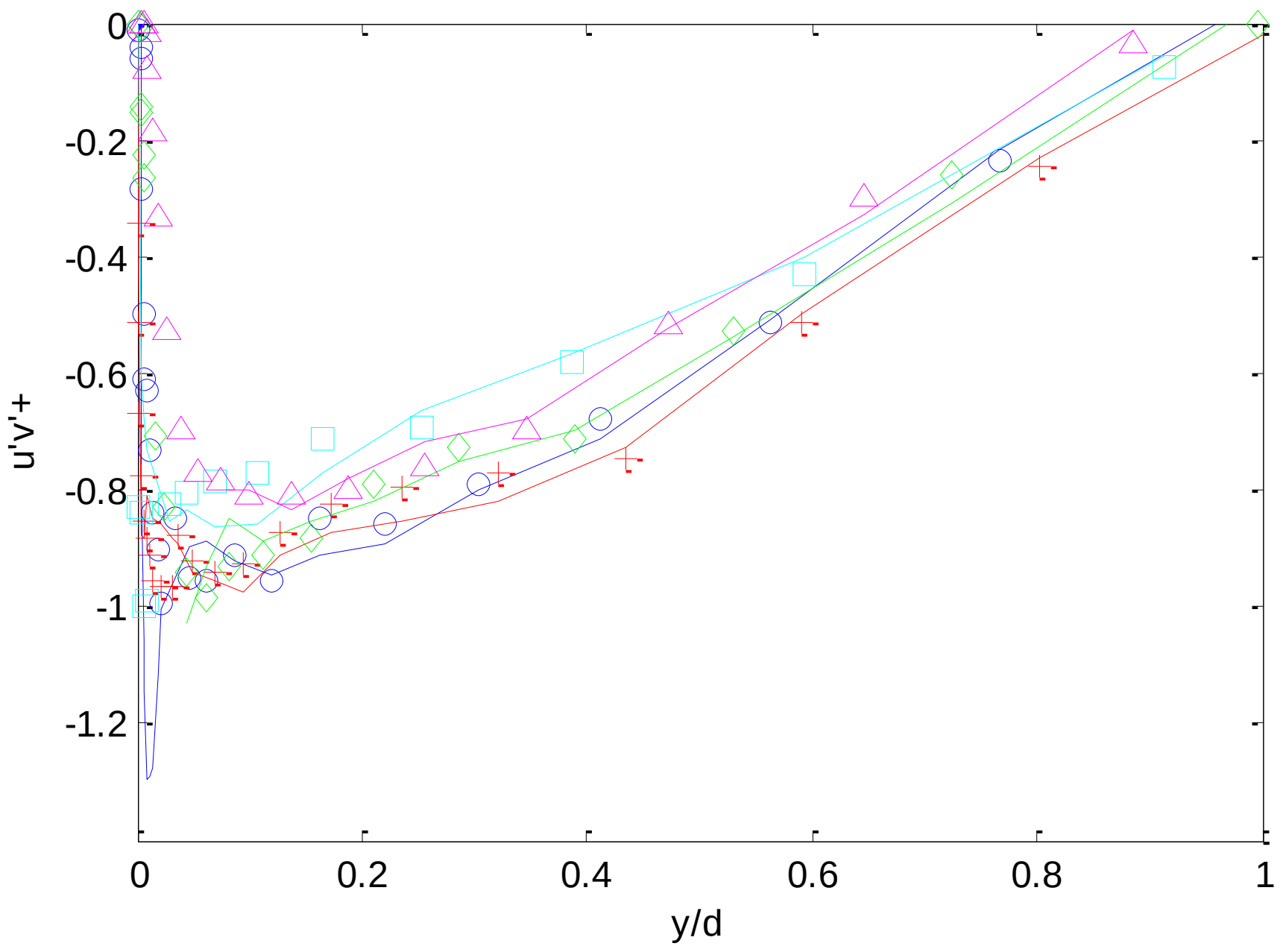

**Fig. 2. Comparison of the Reynolds stress in boundary layer flows over a flat plate. The data (symbols) are for $Re_\theta$ = 1430 ~ 31000 [8]. Lines represent current result (Eq. 1).**

u'$^2$ profiles have a sharp negative peak, near the wall, and gradually decrease as the distance from the wall increases [8, 9]. Although this is somewhat attenuated by the mean velocity term in Eq. 1 (the first term on the RHS of Eq. 4), the resulting contribution of u'$^2$ is still high near the wall. However, the sharp peaks in the u'$^2$ are also associated with a large gradient in u'$^2$, which increases the magnitude of the viscous "dissipation" term, offsetting the large "transport" effect. As noted above, dealing with experimental data does lead to some errors, and some "leakage" in the Reynolds stress is found in Figure 1. Unless the gradients are accurately entered as input, and their y coordinates are perfectly aligned, this kind of leakage can occur. Also, the method is not able to track the Reynolds stress at the lowest Reynolds number close to the wall for the same reason, which is somewhat surprising as the requirements for spatial and measurement resolutions are the least at that condition.

To circumvent the difficulty of using discontinuous data and also to examine the contributing terms from the RHS of Eq. 1, we can fit a simple function to find a smooth, continuous curve for $u'^2$ and U. As an example, we use the data for $Re_\theta$ = 13100 from DeGraaf and Eaton [8], and use $u'^2(y) = A\left[1-\left(\frac{y}{\delta}\right)^{\frac{1}{m}}\right]+B$, which seems to follow the data reasonably well as shown in Figure 3(a) with m = 28. For the mean velocity, we use $(U/U_e) = (y/\delta)^{1/n}$, with n = 7, also shown in Figure 3(a). Then, these parameters can be algebraically or numerically differentiated to input their respective derivatives into Eq. 1, to calculate the Reynolds stress. Although the fit is not quite perfect, it is actually quite functional in showing how the resolution and alignment of the data for root parameters affect the final accuracy of the Reynolds stress calculation near the wall. For example, the curve fit underestimates the steep descent rate of $u'^2$ near the wall in Figure 3(a), but the resulting Reynolds stress is also underestimated at the same location as shown in Figure 3(b). The agreement with experimental data in Figure 3(b) is quite good, and no "leakages" or anomalous peaks in the Reynolds stress are found, in spite of the sharp peak and slope near the wall. The important point of this exercise is that if smooth, continuous, and mutually aligned data are used in Eq. 1, then the Reynolds stress is almost exactly calculated. Therefore, the mean velocity (low near the wall) and the viscous term (high but with an opposite sign) modulate the peaks in $u'^2$ profile near the wall, resulting in the correct prediction of the Reynolds stress based on these basic turbulence parameters.

Figure 3 also illustrates the relatively simple physics of Reynolds stress production and distribution. The leading term on the RHS of Eq. 1 is the dominant term, and is modulated by the viscous term near the wall where again the gradient is extremely high. The integral term reduces the magnitude of the Reynolds stress, particularly at large y locations, where there would otherwise be some deviations. Thus, the turbulent "momentum" or kinetic energy is transported by the mean velocity (the first term on the RHS of Eq. 1), the integral term "displaces" this excess transport term, and the viscous effect dampens the high transport term in region(s) of steep gradients. Therefore, Eq. 1 contains the "forces and balances" involved in the Reynolds stress production and distribution. It also brings an interesting point where all of the terms in Eq. 1 would go to zero except for the integral term, as y approaches the freestream. As noted above, the integral term is a "displacement" term and outside the boundary layer turbulence quantities are no longer displaced, so the displacement or the integral term should go to zero. That is one way the freestream boundary condition can be applied. This effect is relatively insignificant for flows over a flat plate, since there is a prolonged "trail" of the turbulent kinetic energy ($u'^2$) which extend even slightly beyond $y/\delta = 1$ [8]. Thus, the calculated Reynolds stress starts to overshoot above zero before but very close to $y/\delta = 1$ in Fig. 3. For confined flows like the channel flow, to be discussed next, this becomes more of an issue, because the displacement term must cancel out at the centerline. We discuss a method to apply such boundary conditions in the next section.

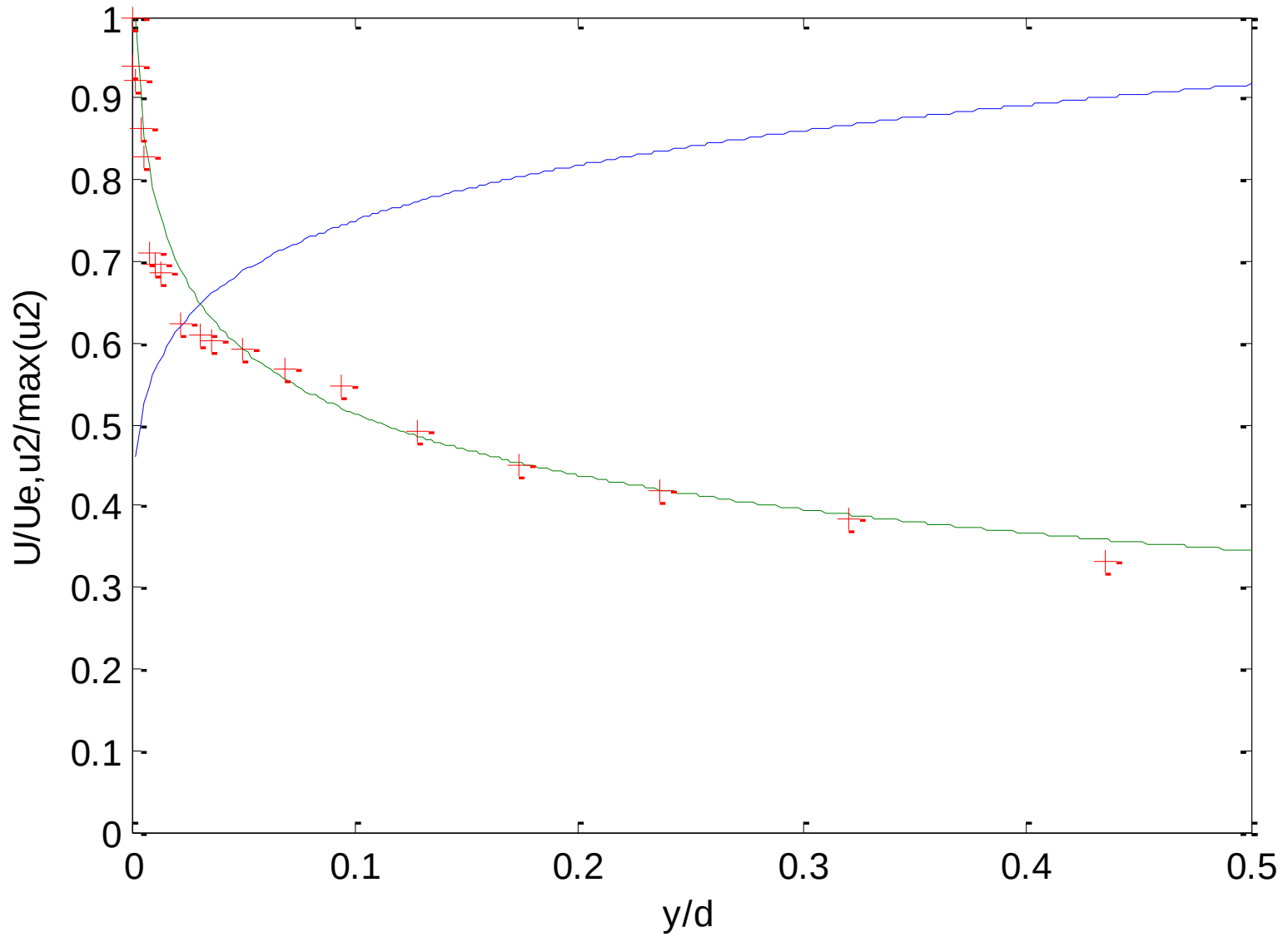

**(a)**

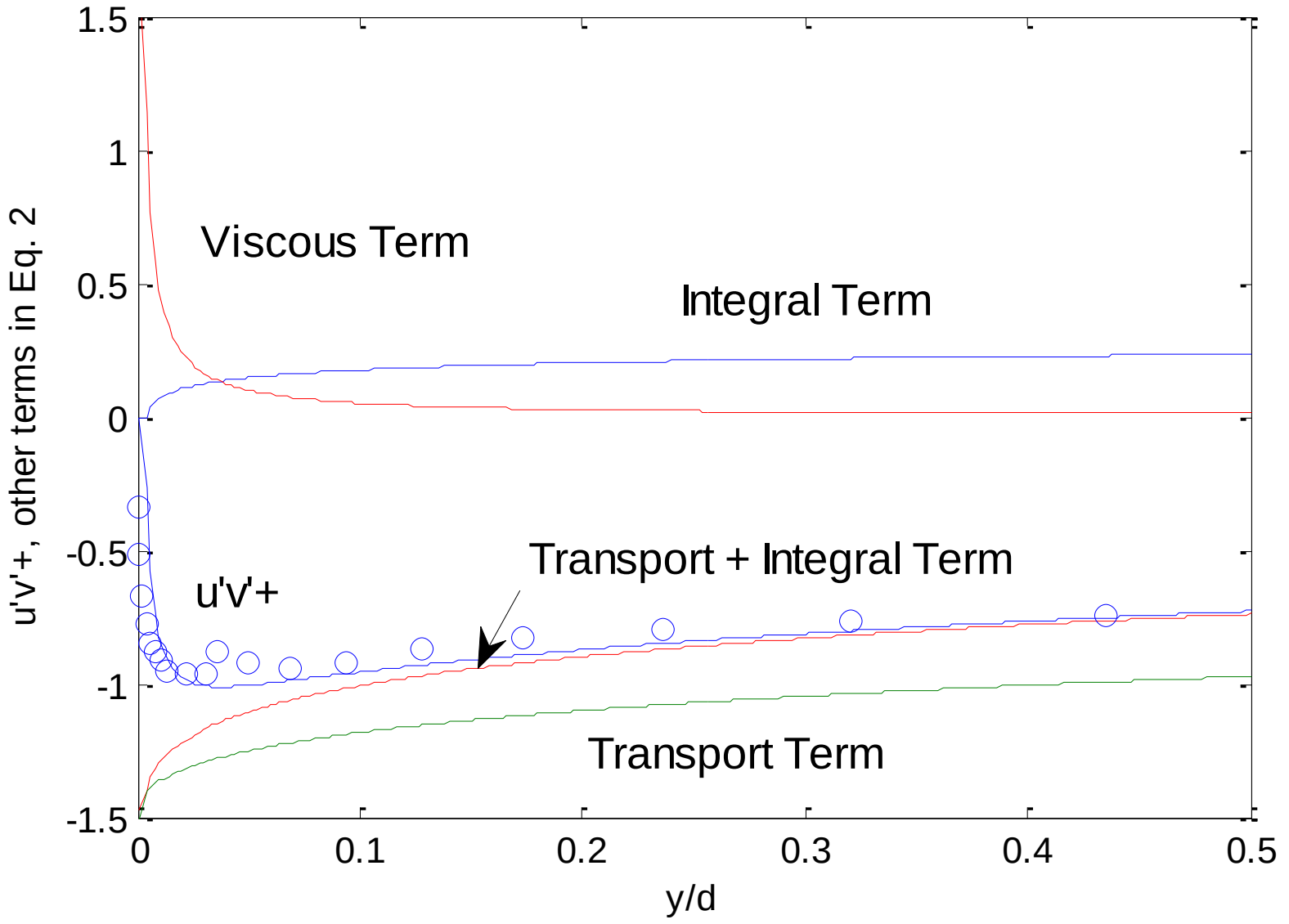

**(b)**

**Fig. 3. Examination of the terms in Eq. 1, based on the input parameters from experimental data [8]. $Re_\theta$ = 13100.**

## IV. Comparison and Validation for Rectagular Channel Flow

Figure 4 is a similar comparison of Reynolds stress as calculated by Eq. 1, with DNS results for fully-developed channel flows [10, 11]. DNS results are highly resolved (we had access to the entire data set through the authors' website [11]), and continuous, and therefore inputting the root parameters and taking the gradients do not lead to much fluctuations or misalignments, as was the case in Figure 2. Iwamoto and co-workers [10, 11] have performed DNS, using established methods for $Re_\tau$ = 110 ~ 650, where $Re_\tau$ is the Reynolds number based on the friction velocity and channel half-width. The entire data set from the DNS is available on their website [11], including the mean velocity, turbulent fluctuating velocity components, and various moments of their products. We input the necessary root turbulence parameters into Eq. 1, and compare with the Reynolds stress from DNS. The agreement is nearly perfect at low Reynolds numbers in Figure 4, which gives some confidence that we have captured the true physics of turbulent transport, and that the results are not fortuitous.

As noted earlier, the constants $C_1$ and c in Eq. 1 vary in a consistent and predictable manner as a function of the Reynolds number, $Re_\tau$. The results in Figure 4 have been obtained with $C_1 = 1.95 \times 10^{-4} (0.352\,\mathrm{Re}_\tau)^{\frac{1}{2.5}}$, and c increases linearly from 0.125 at $Re_\tau$ = 0.125 to 0.425 at $Re_\tau$ = 650. The reason for the increase in these constants is that the displacement effect is amplified at higher Reynolds numbers. The functional dependence of the constants may be deduced in a future work, but in this first study we have used the data to calibrate for the constants as given above.

There is an interesting departure at higher Reynolds numbers, as the solution starts to overshoot the DNS data as y approaches the centerline. The y location where this departure starts to occur decreases (further away from the centerline) at higher Reynolds numbers. This departure is due to the fact that the symmetry boundary condition for channel flows, at the centerline, has not yet been imposed. As noted earlier, the integral term is a “displacement” term accumulating from the wall, and at the centerline the displacement must cancel out. For example, the integral formula (Eq. 1) and its transport equation (Eq. 2) have been derived for flows bounded on one side, such as the flow over a flat plate, and the solution proceeds from y = 0 (wall) onward. For channel flows, the flow is bounded on both sides, leading to the requisite symmetry condition at the centerline. One way to impose the symmetry boundary condition is to force the constant $C_1$ to be proportional to the velocity gradient. For example,

$$C_1 = C_o \left( \frac{\left( \frac{\partial U}{\partial y} \right)}{\left( \frac{\partial U}{\partial y} \right)_{y=0}} \right)^m \qquad (3)$$

With m = 1/3, indeed the calculated Reynolds stress tracks the DNS data fairly well at the Reynolds number of 400, as shown in Figure 5. There are small undulations now that derivatives of mean flow velocity are used in the multiplicative constant. Although this approach may not seem so elegant, Eqs. 1 and 2 have been derived based on the displacement of turbulence parameters, so that symmetry or other boundary conditions should be applied as in Eq. 3. For low Reynolds numbers, the displacement apparently is insignificant and Eq. 3 was not needed, as shown in Figure 4.

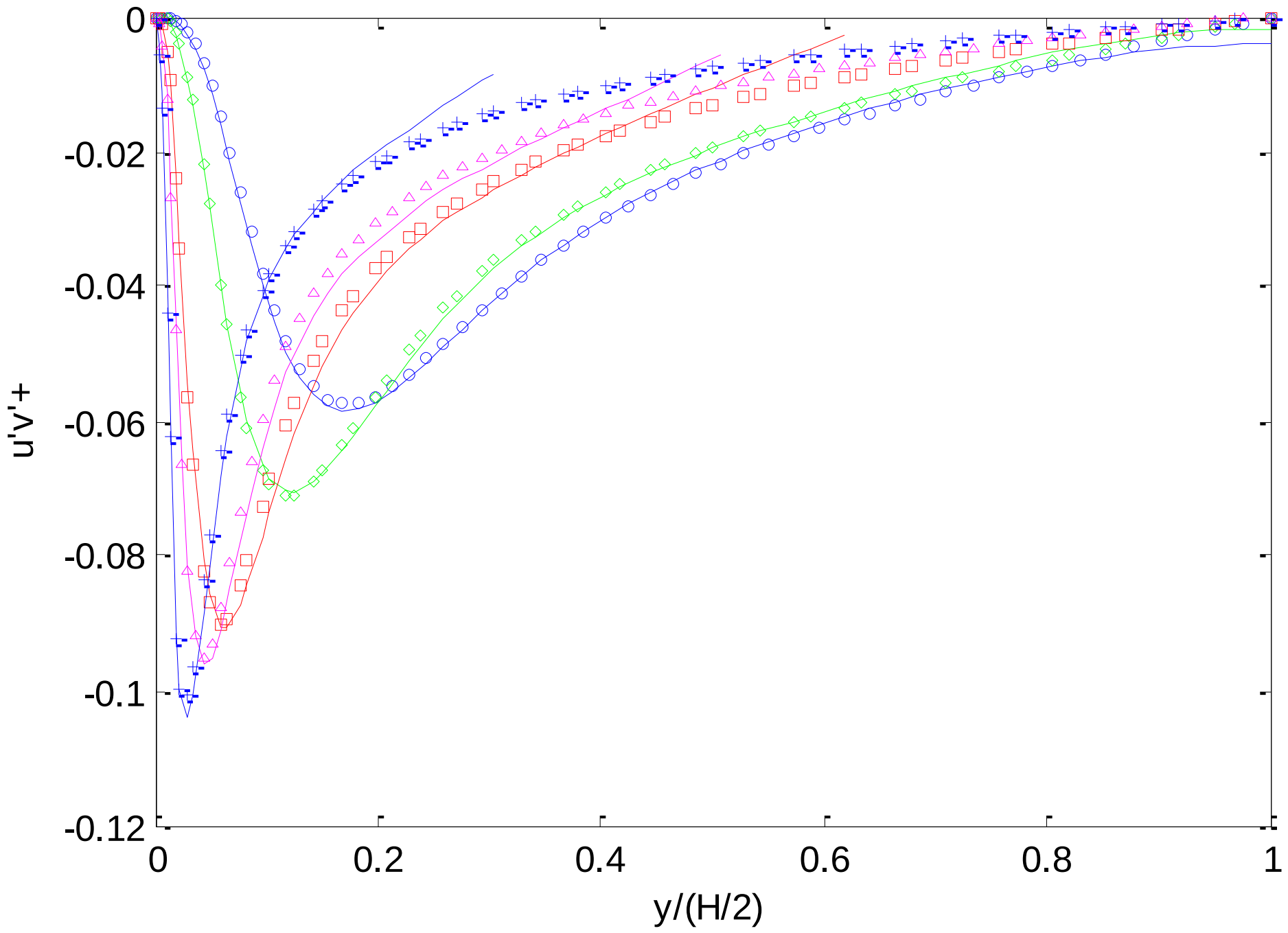

**Fig. 4. Comparison of the Reynolds stress in boundary layer flows over a flat plate. Lines are theoretical results, using Eq. 1. Data symbols: circle ($Re_\tau$ = 110), diamond (150), square (300), triangle (400), + (650) [10].**

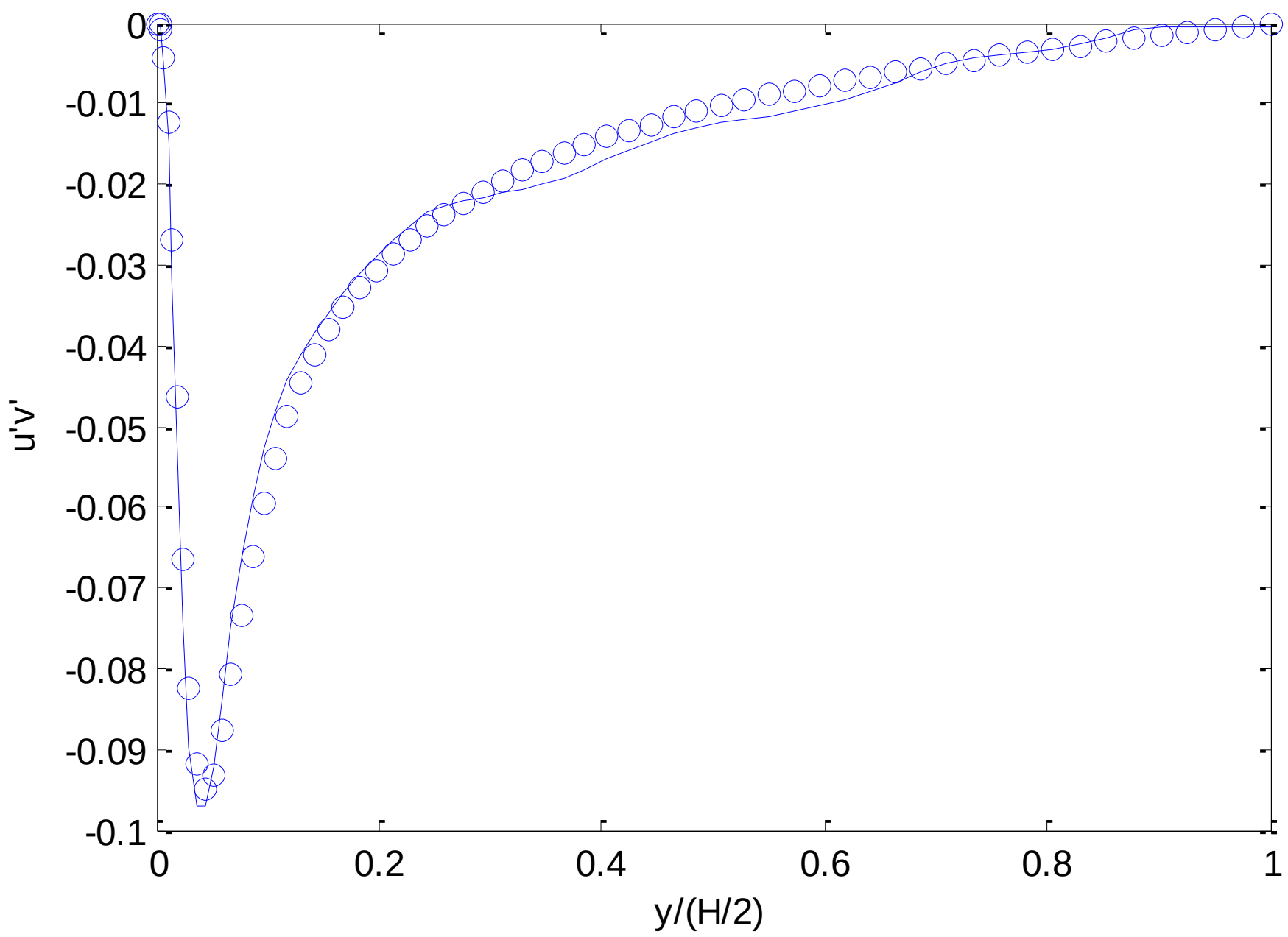

**Fig. 5. Correction for the symmetry condition, using Eq. 5.**

**$Re_\tau$ = 400.**

The development of Eq. 2 (as shown earlier and in the Appendix) represents the conservation of the turbulence momentum, in a moving coordinate frame to isolate the Reynolds stress. We can validate this turbulence momentum balance, along with the differential transform for the boundary layer displacement (Eq. A2 in the Appendix), for the channel flow. Using the transform (Eq. A2), Eq. 2 becomes

$$\frac{\partial(u'v')}{\partial y} = -C_1 U \frac{\partial u'^2}{\partial y} + \nu_m \frac{\partial^2 u_{rms}'}{\partial y^2} \qquad (4)$$

We can verify the validity of this turbulent momentum balance. We first obtain the left-hand side (LHS) of Eq. 4 from u'v' in the DNS data [11], by computing its derivative $\frac{\partial(u'v')}{\partial y}$ numerically. We can then compare with the right-hand side (RHS) of Eq. 4. Only the data for U and $u'^2$ are needed, since $u'_{rms}$ is equal to the square root of $u'^2$. The derivatives are computed using a second-order method. This comparison is shown in Fig. 6, where the LHS and RHS of Eq. 4 are quite well balanced, in spite of the fact that we are taking first- and second-derivatives of discrete data. Therefore, the conservation of turbulence momentum (Eq. 4), along with the differential transform (Eq. A2), holds, in canonical flow geometries. This is also the reason that the Reynolds stress is reproduced by the integral formula (Eq. 1) as shown in Figs. 2-4, because it is based on correct physics of the Reynolds stress.

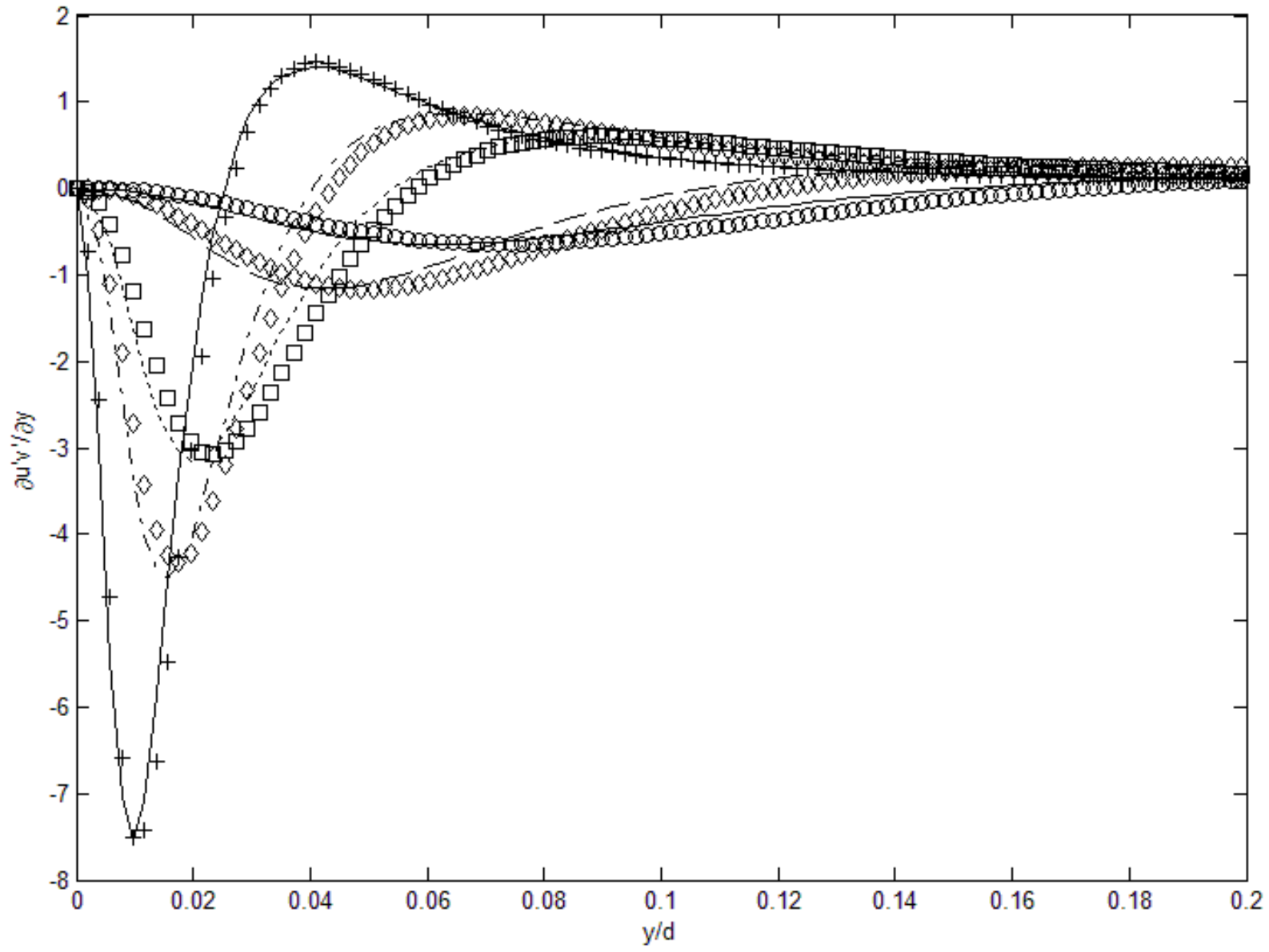

**Fig. 6. Comparison of the LHS and RHS of Eq. 4. Data symbols are LHS of Eq. 4, $\frac{\partial(u'v')}{\partial y}$: circle ($Re_\tau$ = 110), diamond (150), square (300), triangle (400), + (650). Lines represent the RHS of Eq. 4: Transport + Viscous Term**

We can further examine the turbulence momentum balance of Eq. 4, for the channel flow. Each term in Eq. 4 is plotted for $Re_\tau$ = 400, as an example (other Reynolds numbers show the same trend), in Fig. 7. The transport term, $-C_1 U \frac{\partial u'^2}{\partial y}$, has a broad influence on the Reynolds stress, similar to the flow over a flat plate, while the viscous term is significant close to the wall. Near the wall, both of these terms are negative, causing the Reynolds stress to decrease sharply near the wall. The viscous term becomes negligible away from the wall, and the Reynolds stress gradient is mostly determined by the transport term in Eq. 4. This then is the reason that the Reynolds stress and $u'^2$ profiles look similar away from the wall, but the Reynolds stress becomes steep close to the wall. Thus, the current formulation shows that complex modeling of the Reynolds stress is not necessary, and we can use, at least for canonical flows, Eq. 4 to determine the Reynolds stress from root turbulence parameters [12-14]. Only U and $u'^2$ are needed. One can write transport equations for these variables. The U transport equation is simply the Reynolds-averaged momentum equation in the x-direction, and $u'^2$ is derived similar to the turbulent kinetic energy equation. These transport equations are all coupled to the turbulent momentum balance (Eq. 4). In conclusion, Eq. 4 shows the simple physics of how the Reynolds stress is produced/diminished and diffused by the transport and viscous terms, respectively.

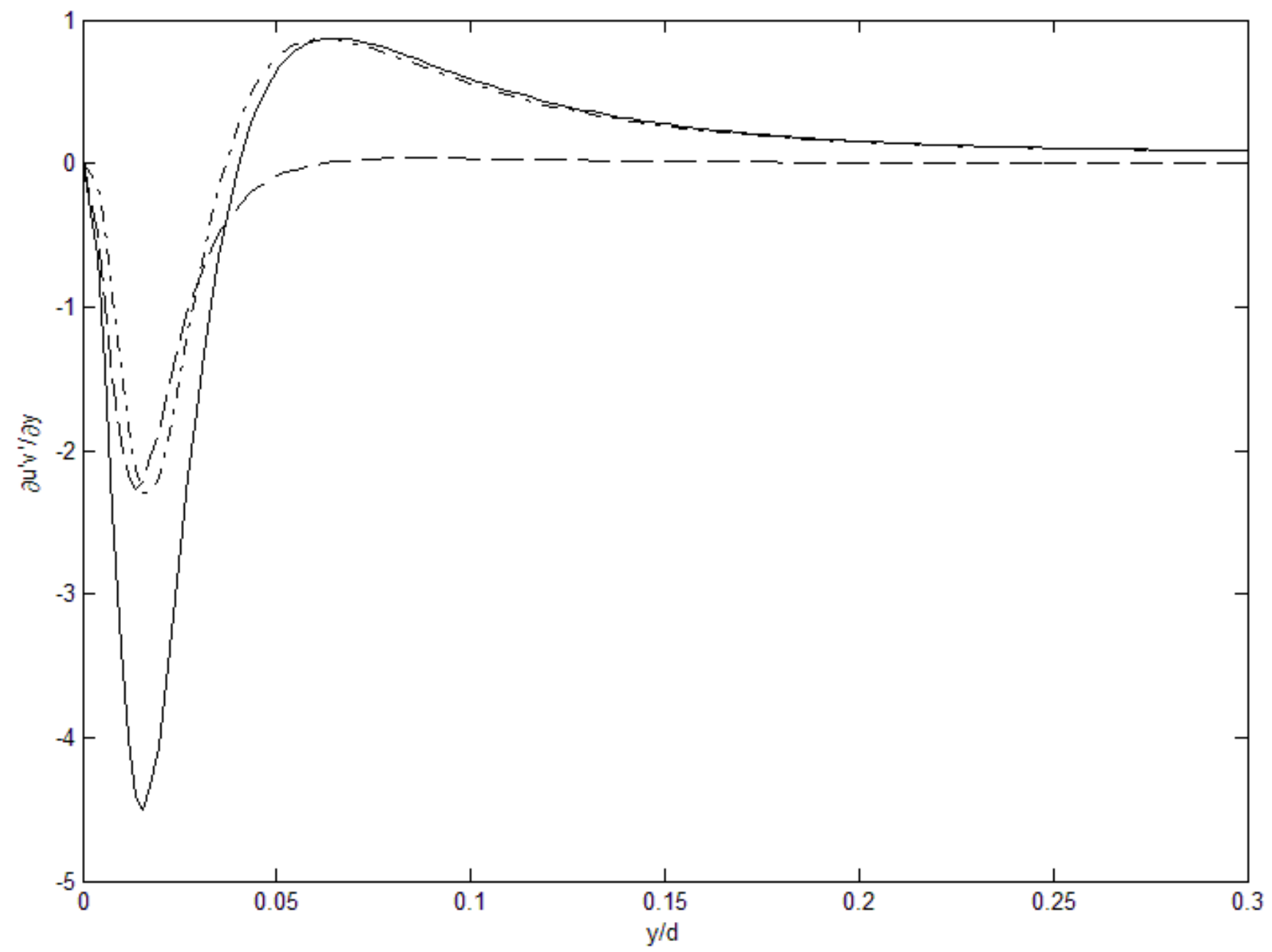

**Fig. 7. Contributing terms to the Reynolds stress gradient, $\frac{\partial(u'v')}{\partial y}$, in Eq. 4. Re$_\tau$ = 400.**

**Solid line: $\frac{\partial(u'v')}{\partial y}$; dash-dot: $-C_1 U \frac{\partial u'^2}{\partial y}$; and dashed: $\nu_m \frac{\partial^2 u_{rms}'}{\partial y^2}$.**

## V. Conclusions

We have found an expression for the Reynolds stress in canonical flows, based on the turbulence momentum balance for a control volume moving at the local mean flow speed. The resulting "integral formula" works quite well in determining the Reynolds stress based on inputs of root turbulence parameters, such as streamwise component of the turbulence kinetic energy, the mean velocity and its gradient. The predicted Reynolds stress is in good agreement with experimental and DNS data. In particular, if the data are continuous and aligned, then agreement is nearly perfect. Using DNS data, the conservation of turbulence momentum (Eq. 4) is shown to be valid, at least in canonical flow geometry, which suggests the possibility for broad applications of this theory. There are some nuances and corrections yet to be examined, such as applying symmetry boundary conditions and relating the displacement effect to the flow geometry and Reynolds number. Also, thus far, the theory has been tested against relatively simple geometries [12, 13], but is extendible to three-dimensional turbulent flows. The displacement effect, for example, will also be three-dimensional in such flows, and schemes for correctly treating this effect are currently being investigated [14].

## Appendix: DERIVATION OF THE INTEGRAL FORMULA (Eq. 1)

In Reynolds-averaged Navier-Stokes (RANS) equation, non-linear terms involving turbulent fluctuation velocities arise since the absolute velocity is decomposed into mean (U) and fluctuating (u') component: $u = U + u'$. The non-linear terms that develop during the averaging process in the RANS are called the Reynolds stress, which involves time-averaged components of products of fluctuating velocities, $u_i' u_j'$. Here, we omit the bar above u'$^2$, u'v', etc, for simplicity, and take all the fluctuation parameters to be time-averaged. Figuring out how the Reynolds stress is related to the mean and other "root" turbulent parameters has been the topic of numerous studies, for quite some time. However, we notice that the decomposition is necessary only in the absolute (stationary) coordinate frame, as shown in the top part of Figure 1. What if we consider a control volume moving at the mean flow? Then, an observer riding this control volume will not "feel" the mean flow, but only the fluctuating components, as shown in the bottom part of Figure 1. We can re-do our momentum balance for this control volume, where only the root-mean-square (rms) fluctuations of the velocities (momentum) need to be balanced, in the x-direction. Momentum balance in other directions would produce other components of the Reynolds stress. The momentum equation is greatly simplified in this relative coordinate frame:

$$\frac{\partial u'^2}{\partial x} + \frac{\partial (u'v')}{\partial y} = \nu \frac{\partial^2 u_{rms}'}{\partial y^2} \qquad \text{(A1)}$$

For stationary flows, the fluctuating momentum gradients are balanced by the viscous force term, and the pressure term is neglected. By using the Lagrangian control volume, the mean momentum terms are de-coupled from the fluctuating terms. We also apply the boundarly layer approximation (V<<U) and neglect the laminar viscous term based on gradient of U. Again, all the quantities ($u_{rms}$', u'$^2$, u'v', and $p_{rms}$') in Eq. A1 are time-averaged. The idea of time averaging in this framework is relatively straight-forward for all of the terms, except for the viscous term. The viscous shear stress is caused by instantaneous gradient of the u' in the y-direction. Thus, the concept here is that there can be some time-averaged gradient (and second gradient) in u' that would then cause the viscous shear stress.

In conventional calculations in the absolute coordinate frame, the x-derivatives would have been set to zero for fully-developed flows, and we would be left with a triviality. However, for a boundary-layer flow as an example (Figure A1), the boundary layer grows due to the "displacement" effect. The mass is displaced due to the fluid slowing down at the wall, as is the momentum, and turbulence parameters as well. The boundary layer thickness grows at a predictable rate, depending on the Reynolds number. Thus, if one rides with the fluid moving at the mean velocity, one would see a change in the all of the turbulence properties, as illustrated in Figure A1, and therefore the term "displaced control volume method". This displacement effect can be mathematically expressed as:

$$\frac{\partial}{\partial x} = C_1 U \frac{\partial}{\partial y} \tag{A2}$$

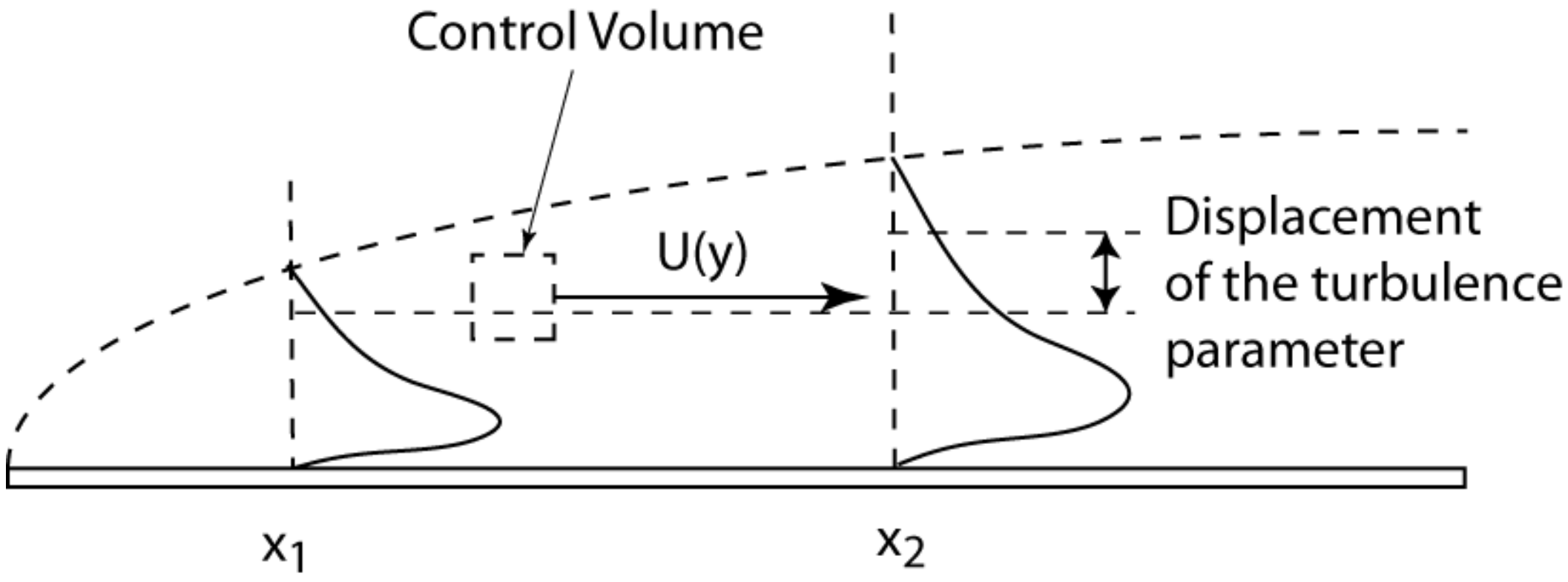

**Fig. A1. A schematic illustration of the concept of the displaced control volume method.**

I.e., the fluid parcel will see a different portion of the boundary layer in the y-direction, and how much difference it will see depends on how fast the fluid is moving along in the boundary layer. Thus, the mean velocity, U, appears as a multiplicative factor in Eq. A2. $C_1$ is a constant that depends on the Reynolds number. Similarly, the gradient of the pressure fluctuation will not be zero in general. However, this term is expected to be significant only for compressible flows, so we omit this term from further analysis in this phase of the work. In Eq. A1, we now have a simple integrable expression to find the Reynolds stress, after using Eq. A2. If we integrate by parts, we obtain:

$$u'v' = -C_1\left[Uu'^2 - c\int_0^y \frac{dU}{dy}u'^2 dy\right] + \nu_m \frac{\partial u_{rms}'}{\partial y} \qquad \text{(A3)}$$